\documentclass[twocolumn,showpacs,preprintnumbers,amsmath,amssymb,floatfix]{revtex4}

\usepackage{graphicx}
\usepackage{dcolumn}
\usepackage{bm}
\begin{document}
 
\title{Shear yielding of amorphous glassy solids: Effect of
temperature and strain rate}

\author{J\"org Rottler\footnote{present address: Laboratoire de Physico-Chimie Th\'eorique, \\ ESPCI, 10 rue Vauquelin, F-75231 Paris Cedex 05, France}}
\email{Joerg.Rottler@jhu.edu}
\author{Mark~O.~Robbins}

\affiliation{Department of Physics and Astronomy, The Johns Hopkins
University, 3400 N.~Charles Street, Baltimore, Maryland 21218}

\date{\today}

\begin{abstract} 
We study shear yielding and steady state flow of glassy materials with
molecular dynamics simulations of two standard models: amorphous
polymers and bidisperse Lennard-Jones glasses. For a fixed strain
rate, the maximum shear yield stress and the steady state flow stress
in simple shear both drop linearly with increasing temperature.  The
dependence on strain rate can be described by a either a logarithm or
a power-law added to a constant. In marked contrast to predictions of
traditional thermal activation models, the rate dependence is nearly
independent of temperature. The relation to more recent models of
plastic deformation and glassy rheology is discussed, and the dynamics
of particles and stress in small regions is examined in light of these
findings.
\end{abstract}


\maketitle

\section{Introduction}
Deformation processes and plasticity in amorphous materials such as
metallic or polymeric glasses have recently received a lot of
attention
\cite{Larson1999,Rottler2001,STZ,Lemaitre2001,Hasan1995}. These
materials are used in many load-bearing applications. However, an
understanding of their yield and flow properties is hampered by the
absence of long-range order and easily identifiable mechanisms that
mediate the deformation such as dislocation motion in crystals. A
similar situation is encountered in ``soft'' glassy materials such as
foams, pastes, and colloidal suspensions, which are also characterised
by a liquid-like structure and long relaxation times. In fact, it has
been suggested recently that these very different materials can be
viewed as particular realizations of a jammed state \cite{Liu2001},
which implies that their mechanical behavior could be described in a
common framework.

Much insight into the mechanical behavior of structural glasses has
been gained from molecular simulations of simple glass-forming liquids
or polymers, where particles interact through a Lennard-Jones
potential. For example, Falk and Langer \cite{STZ} studied
two-dimensional shear deformation of a mixture of such particles and
found localization of plastic events in so-called shear transformation
zones. Barrat and Berthier \cite{Barrat2000} studied the steady state
flow of a similar model and analyzed its relation to the
fluctuation-dissipation theorem in out-of-equilibrium situations.

In a recent paper \cite{Rottler2001}, we have studied the onset of
shear yielding of amorphous polymer glasses under multiaxial loading
conditions. A pressure-modified von Mises criterion \cite{Ward1983}
accurately describes the maximum shear yield stress as a function of
the applied stress for different temperatures.  However, in these
simulations the bulk polymer was deformed at a single constant strain
rate. The present work investigates the effect of strain rate on the
stress at the onset of shear yielding as well as on the steady state
flow stress of glassy materials in simple shear. In order to relate
the onset of yield to steady shear, we not only discuss the
macroscopic material response, but also perform an analysis of the
localized dynamics of the stress distribution in the deforming solid.
Such a program is particulary suited to test predictions of models of
plasticity \cite{Eyring1936,STZ,Lemaitre2001,SGR,Berthier2000} and may
lead to a deeper understanding of the nature of plastic deformation.

In the following section, we briefly summarize traditional and more
recent models of viscoplasticity and rheology of glassy
materials. Sections \ref{model-sec} and \ref{results-sec} discuss the
molecular models and the simulation results, respectively. Section
\ref{disc-sec} critically reviews how the theoretical ideas of Section
\ref{theo-sec} describe the simulation results.

\section{Models of plasticity and rheology}
\label{theo-sec}

\subsection{Rate and temperature dependence: Eyring model}
The simplest model that makes a prediction for the rate and
temperature dependence of shear yielding is the rate-state Eyring
model \cite{Larson1999,Eyring1936} of stress-biased thermal
activation.  Structural rearrangement is associated with a single
energy barrier $E$ that is lowered or increased linearly by an applied
stress $\sigma$.  This defines transition rates of the form
\begin{equation}
R_\pm=\nu_0\exp{\left[-\frac{E}{k_BT}\right]}\exp{\left[\pm\frac{\sigma
V^*}{k_BT}\right]},
\end{equation}
where $\nu_0$ is an attempt frequency and $V^*$ is a constant called
the ``activation volume.'' In glasses, the transition rates are
negligible at zero stress. Thus at finite stress one needs to consider
only the rate $R_+$ of transitions in the direction aided by
stress. The plastic strain rate $\dot{\epsilon}_{pl}$ will be
proportional to $R_+$, $\dot{\epsilon}_{pl}=cR_+$.  Solving for the
stress $\sigma$, one obtains
\begin{equation}
\sigma=\frac{E}{V^*}+\frac{k_BT}{V^*}\ln\left[\frac{\dot{\epsilon}_{pl}}{c\nu_0}\right].
\label{eyring-eq}
\end{equation}

Eq.~(\ref{eyring-eq}) contains only a single relaxation time scale and
predicts an apparent yield stress that varies logarithmically with the
strain rate and a prefactor that depends linearly on
temperature. Despite its simplicity, experimental results
\cite{Ward1983} are often fitted to Eq.~(\ref{eyring-eq}), and the
value of $V^*$ is associated with a typical volume required for a
molecular shear rearrangement.

\subsection{Modern approaches}
Modern phenomenological approaches pay tribute to the complexity of
glassy systems through several extensions.  First, it has been
realized that assuming a single energy barrier for rearrangements is
an oversimplified description of glassy materials
\cite{Hasan1995}. One can therefore introduce a distribution of
barriers and add additional time scales. Second, any theory that
attempts to predict a full stress-strain curve must contain some
information about the internal state of the system as a function of
time or strain. Extensions therefore consider dynamical internal state
variables. In the following, we describe particular realizations of
these ideas.

\subsubsection{Shear transformation zone theory}
Falk and Langer used molecular dynamics simulations very similar to
the present study to identify local plastic rearrangements under
shear. They formulated a theory of viscoplasticity \cite{STZ} based on
the concept of ``shear transformation zones'' (STZ), bistable
(mesoscopic) regions that transform under shear between
$\pm$-states. One then considers the dynamics of an ensemble of STZ
with number density $n_{\pm}$ on a mean-field level, which determines
the plastic strain rate
\begin{equation}
\dot{\epsilon}_{pl}=A_0(R_+n_+-R_-n_-),
\end{equation}
where $A_0$ is a constant. A key difference from the Eyring model
resides in the form of the transition rates $R_\pm$, which are assumed
to be free-volume (entropically) activated rather than thermally
activated, i.e.
\begin{equation}
R_\pm=R_0\exp{\left[-\frac{v_0\exp{[\mp\sigma/\bar{\mu}]}}{v_f}\right]},
\end{equation}
where $v_0$ is a characteristic free volume required for a STZ flip and
$\bar{\mu}$ a characteristic stress scale required for a molecular
rearrangement. The role of temperature is played by a ``free volume''
$v_f$ per particle.  The authors motivate this with the observation
that in a solid at very low temperature, energy barriers should be
very large compared to thermal energies and thus, as in granular
systems, thermal activation over these barriers should be
negligible. The population densities themselves evolve according to
the rate equation
\begin{equation}
\dot{n}_\pm=R_\mp
n_\mp-R_{\pm}n_{\pm}+\sigma\dot{\epsilon}_{pl}(A_c-A_an_\pm),
\end{equation}
where the last term introduces creation and annihilation processes of
STZ's proportional to the work of plastic deformation
$\sigma\dot{\epsilon}_{pl}$. The STZ equations can be solved
analytically in special steady state situations and otherwise solved
numerically. They were shown \cite{STZ} to have both a jammed solution
for which $\dot{\epsilon}_{pl}=0$ and a flowing solution once $\sigma$
exceeds a true yield stress $\sigma_y$. The shear rate rises linearly
as sigma increases above $\sigma_y$ as in a Bingham fluid.  Numerical
stress-strain curves, hysteresis experiments and creep tests were
shown to be accurately reproduced by the model after the adjustment of
several fit parameters.

Lema\^\i tre \cite{Lemaitre2001} recently extended STZ theory with
concepts from the physics of granular media. He treated the free
volume parameter $v_f$ as a dynamical state variable and proposed the
following time evolution:
\begin{equation}
\label{volrelax-eq}
\dot{v}_f=-R_1\exp{\left[-\frac{v_1}{v_f}\right]}+A_v\sigma\dot{\epsilon}_{pl}.
\end{equation}
This expression is motivated by slow density relaxations in granular
materials that decrease the free volume. The activation factor
$\exp{[-v_1/v_f]}$ describes the probability for volume fluctuations
larger than a characteristic volume $v_1$. Note that the ``activation
barrier'' for compaction $v_1$ differs {\it a priori} from the barrier
for shear transformation $v_0$. The second term refers to creation of
free volume (shear induced dilatancy) again due to plastic
deformation. A "linearized" version of this theory produces a power
law relation between shear rate and shear stress, $\sigma\sim
\dot{\epsilon_{pl}}^{\kappa-1/\kappa+1}$, where $\kappa=v_1/v_0$.  The
full theory can yield more complicated functional forms with a true
yield stress \cite{Lemaitre2001}.

\subsubsection{Soft Glassy Rheology model}
The soft glassy rheology (SGR) model of Sollich {\it et al.}
\cite{SGR} is an extension of a trap model for glasses originally
proposed by J.-P.~Bouchaud \cite{Monthus1996} with stress acting as an
external drive. It was designed to describe the flow behavior of
foams, dense emulsions, pastes and slurries.  However, it is very
similar to the previous models and should also be relevant to the
materials of interest here. Small volume elements are assumed to yield
with a rate $\Gamma_0\exp[-(E-kl^2/2)/x]$, where $l$ is the local
strain and $k$ an elastic constant. The role of temperature is
replaced by a ``noise'' temperature $x$, which is assumed to describe
the effect of structural rearrangements in a mean-field spirit. Note
that the model describes stress-assisted yielding as in the other two
models, but stress enters quadratically and not linearly. Structural
disorder is modeled with an exponential distribution of yield energies
$E$, $\rho(E)=\exp(-E/x_g)/x_g$, where $x_g$ is usually set to
one. This introduces a distribution of relaxation timescales.  After
yielding of a volume element, a new yield energy is drawn from
$\rho(E)$ and the local strain $l$ rises again from 0 according to a
macroscopic shear rate $\dot{\gamma}$.  The time evolution of the
probability $P(l,E;t)$ describing the ensemble of volume elements can
be obtained from a master equation.

Like the STZ theory, the predictions arising from the SGR equations
are richer than the simple Eyring model. The exponential distribution
of traps induces a dynamical glass transition, and the system exhibits
aging for $x<1$. Analysis has mainly focused on the steady state
situation under constant shear rate $\dot{\gamma}$, which is the
generic experiment used to determine the mechanical properties of soft
glassy materials. Salient predictions are: a Newtonian fluid flow
$\sigma\propto \dot{\gamma}$ for $x>2$ and a power law fluid
$\sigma\propto \dot{\gamma}^{x-1}$ for $1<x<2$ \cite{SGR}. In the
glassy phase, a scaling of the form
$\sigma-\sigma_y\propto\dot{\gamma}^{1-x}$ is predicted.  The nature
of the noise temperature $x$ (not the true thermodynamic temperature)
and the pre-exponential factor $\Gamma_0$ (thermal) remain largely
unspecified. 

\subsubsection{Microscopic Approaches}
The descriptions of nonlinear rheology and plasticity described so far
are appealing because of their simplicity, but much of the physics is
put in ``by hand''. In the most recent literature, efforts are made to
derive the response of driven glassy systems from microscopic
considerations. Berthier {\it et al.} \cite{Berthier2000} consider a
generic driven glassy system by calculating the correlation and
response functions from the microscopic Langevin equations in a
mean-field approximation. Based on this approach, they suggest a
``two-time-scale scenario,'' in which the slow time scales associated
with structural relaxations are accelerated by the drive, while the
fast degrees of freedom (phonons) remain at the thermodynamic
temperature.  This concept can be extended to introduce an ``effective
temperature'' different from the thermodynamic temperature. A relevant
conclusion for the present work is that in their calculations for
$T>T_c$ \cite{tc-comm}, the slow relaxation time $t_\alpha$ decreases
with the drive, $t_\alpha\sim\dot{\gamma}^{-2/3}$. Since the
relaxation time determines the viscosity, this leads to power law
shear thinning, $\sigma\sim \dot{\gamma}^n$ with exponent
$n=1/3$. Below $T_c$, their numerical results indicate that $n$ is
only very weakly temperature dependent.

\section{Molecular Simulations}
\label{model-sec}
We perform three-dimensional molecular dynamics (MD) simulations for
two model glasses. A bead-spring model is used for polymers. Beads of
mass $m$ interact via a conventional 6-12 Lennard-Jones (LJ)
potential. All results will be expressed in terms of the
characteristic length $a$, energy $u_0$ and time $\tau_{\rm
LJ}=\sqrt{ma^2/u_0}$ of this potential. Unless otherwise noted, the
potential is truncated at $r=r_c=1.5a$ for computational
convenience. We construct linear polymer chains by connecting adjacent
beads with the finite extensible nonlinear elastic (FENE) bond
potential \cite{Kremer1990}. This polymer model has been used
extensively to study polymer melt dynamics \cite{Puetz2000} and was
also used in our previous study of yield conditions
\cite{Rottler2001}. The number of beads per chain used below is
usually 256, but the chain length and entanglement effects are not
important for the small strains considered in Section \ref{onset-subsec}.

In addition to the polymer, a binary mixture composed of 80\%
A-particles and 20\% B-particles without covalent bonds is also
studied. LJ interaction parameters were set to the values employed in
previous studies that aimed at verifying predictions of mode-coupling
theory for supercooled liquids \cite{Kob1995}, and studied aging
\cite{Kob1997} or dynamical heterogeneities \cite{Kob1997b} during the
glass transition. A very similar system with slightly different
parameters was used by Falk \cite{STZ} to study deformation and
plasticity in amorphous metals in two dimensions.

Both the polymer and binary mixture models enter an amorphous glassy
state without crystallization upon cooling. For the polymer model, the
glass transition temperature $T_g\approx 0.35 \pm 0.05 u_0/k_B$, while
$T_g$ is smaller for the binary system. These values are for $r_c=1.5a$,
and slightly higher values of $T_g$ are obtained with larger $r_c$
\cite{Bennemann1998}.  Here, we focus on a temperature range between
$T=0.01u_0/k_B$ and $T=0.3u_0/k_B$. Unless noted, the temperature is
controlled with a Langevin thermostat (damping rate $1\tau_{\rm
LJ}^{-1}$). The equations of motion are solved using the velocity
Verlet algorithm with a timestep of $dt=0.0075\,\tau_{\rm LJ}$.

The simulation cell contained 32768 LJ beads. Previous studies of the
yield behavior \cite{Rottler2001} have shown that an increase of the
system size beyond this size leads to slightly lower values of the
shear yield stress, but the generic behavior is unchanged. In order to
minimize statistical fluctuations while varying the strain rate, we
use the same initial state for all runs at a given temperature. The
glassy states were prepared by a quench from a fluid temperature
of $T=1.3 u_0/k_B$ to a glassy temperature of $T=0.3 u_0/k_B$ at
constant volume over a time interval of order 1000$\tau_{\rm LJ}$.
The density was chosen so that the hydrostatic pressure was zero at
this temperature. Lower temperatures were then reached by cooling at
constant pressure.

In studies of initial yield behavior, periodic boundary conditions are
applied in all directions to eliminate edge effects. The original cell
is cubic and at zero hydrostatic pressure. Tensile or compressive
strains are imposed on one or more axes by rescaling the periods
$L_i$. We use true strain rates $\dot{\epsilon}_i=L_i^{-1}dL_i/dt$
between $\dot{\epsilon}_i= 10^{-6}\,\tau_{\rm LJ}^{-1}$ and
$\dot{\epsilon}_i= 10^{-3}\,\tau_{\rm LJ}^{-1}$.  Note that since
$\tau_{\rm LJ}\sim 3\,{\rm ps}$, the strain rates employed are much
higher than typical experimental strain rates. As discussed below,
sound propagation is too slow for stress to equilibrate across the
system at the highest strain rates. However, most simulations are slow
enough that loading proceeds nearly quasistatically.

Steady state flow cannot be investigated by deforming the simulation
box in the above manner, because some box periods would soon decrease
to a single molecular diameter. One approach is to use Lees-Edwards
boundary conditions \cite{Allen1987}. Here, we choose a different
route and replace the periodic boundary conditions in the 3-direction
with two rigid walls composed of 2 layers of an fcc(111) crystal for
simulations of simple shear. Otherwise, the simulation cell has the
same dimensions and size, and the wall beads are strongly coupled to
the sheared glass so that slip at the interface is prohibited. By
moving one wall parallel to the solid at constant velocity $v_x$, a
steady state shear profile is imposed, and we can measure the shear
stress $\tau$ as a function of the average strain $\gamma=tv_x/h$,
where $h=32a$ is the wall separation and $t$ is the elapsed time. In
these simulations, the Langevin thermostat is only coupled to the
irrelevant (y) direction perpedicular to the flow (x) and velocity
gradient (z) directions \cite{Thomson1990}.

\section{Results}
\label{results-sec}
\subsection{Onset of shear}
\label{onset-subsec}
\begin{figure}[t]
\includegraphics[width=8cm]{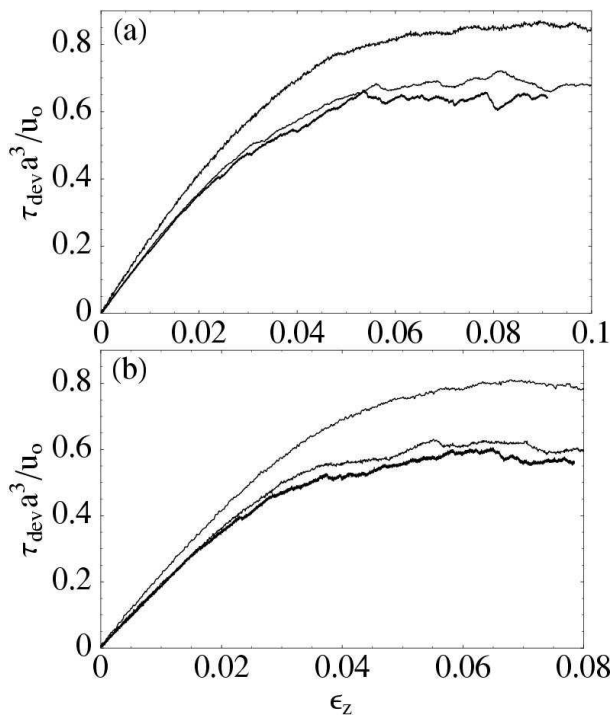}
\caption[Stress-strain curves for different strain rates]
{Stress-strain curves for (a) the polymer glass and (b) the 80/20 LJ
glass at $T=0.01 u_0/k_B$ and three different strain rates
$10^{-3}\tau_{\rm LJ}^{-1}$, $10^{-4}\tau_{\rm LJ}^{-1}$, and
$10^{-5}\tau_{\rm LJ}^{-1}$ (from highest to lowest curves).}
\label{stresstraindiffrates-fig}
\end{figure}

In ref.~\cite{Rottler2001} we analyzed the onset of shear deformation
in polymer glasses.  Results for a wide range of multiaxial loading
conditions, temperatures and potential parameters were consistent with
the pressure-modified von Mises criterion.  In this criterion, the
driving force for shear is the deviatoric stress $\tau_{\rm
dev}=\left((\sigma_{1}-\sigma_{2})^2+(\sigma_{2}-\sigma_{3})^2+(\sigma_{3}-\sigma_{1})^2\right)^{1/2}/3$,
where the $\sigma_i$ are the three eigenvalues of the stress tensor.
Rather than having a sudden onset at a finite strain, irrecoverable
deformation was observed at arbitrarily small strains.  The most
robust definition of the yield stress was $\tau^y_{\rm dev}$, the peak
value of the deviatoric stress as a function of strain.  This quantity
increases linearly with pressure in agreement with the
pressure-modified von Mises criterion.

The results in ref.~\cite{Rottler2001} were obtained for a single
value of the strain rate.  Here we focus on the effect of varying
strain rate on the shear yield stress.  The initially cubic simulation
cell is expanded in one direction at a constant strain rate
$\dot\epsilon_z = L_z^{-1} dL_z/dt$, and volume $V$ is conserved by
maintaining $L_x =L_y =\sqrt{V/L_z}$.
Fig.~\ref{stresstraindiffrates-fig} shows the deviatoric stress
averaged over the entire simulation cell as a function of strain at
three different strain rates.  As can be seen, the behavior of the
bead spring model (a) and the binary LJ glass (b) is qualitatively
similar.  These curves also closely resemble experimental
stress-strain curves for e.~g.~polymeric glasses \cite{Hasan1995}.

For all systems the initial response is nearly elastic, i.e. the
stress rises almost linearly with strain. In the quasistatic limit,
the initial slope gives the elastic modulus. Results for strain rates
of $3\times 10^{-4}\tau_{\rm LJ}^{-1}$ and below collapse onto this
quasistatic behavior.  As the strain rate rises to $10^{-3}\tau_{\rm
LJ}^{-1}$ and beyond, the initial slope grows. The reason is that
stress is no longer able to equilibrate across the system. It is well
known that the elastic modulus of a heterogeneous system is
over-estimated by applying a uniform strain (the Voight limit
\cite{Voight1928}).  Our algorithm imposes a uniform strain at each
step by rescaling the cell dimensions and particle coordinates and the
resulting stress will be too high when the strain rate becomes too
large. A rough estimate for the characteristic time for stress
equilibration through the system is $L/c \sim 10 \tau_{\rm LJ}$, where
$c$ is the speed of sound.  When the strain is stopped suddenly we
find an exponential stress relaxation with a characteristic time $\sim
15 \tau_{\rm LJ}$.  At a strain rate of $\dot{\epsilon_i}=
10^{-3}\tau_{\rm LJ}^{-1}$, the time to reach a strain of 2\% is $20
\tau_{\rm LJ}$ and becomes comparable to the above estimates. We
conclude from Fig.~\ref{stresstraindiffrates-fig} that the shear rate
is slow enough to produce a quasistatic deformation for
$\dot{\epsilon_i} < 10^{-3}\tau_{\rm LJ}^{-1}$, but not at the highest
shear rates.

As the strain increases, even the curves for lower strain rates split
apart and saturate at different maximum heights $\tau_{\rm dev}^y$.
The maxima are broad and centered at strains of about 6\% in the
binary LJ glass and 8\% in the polymer.  The fine structure on the
curves corresponds to individual plastic yield events that are
discussed further below.  This structure decreases with increasing
system size and temperature as the fraction of the system involved
with typical events decreases.  Results for larger systems were
consistent with those shown here, but could not be extended over as
wide a range of shear rates and other parameters.

\begin{figure}[t]
\begin{center}
\includegraphics[width=8cm]{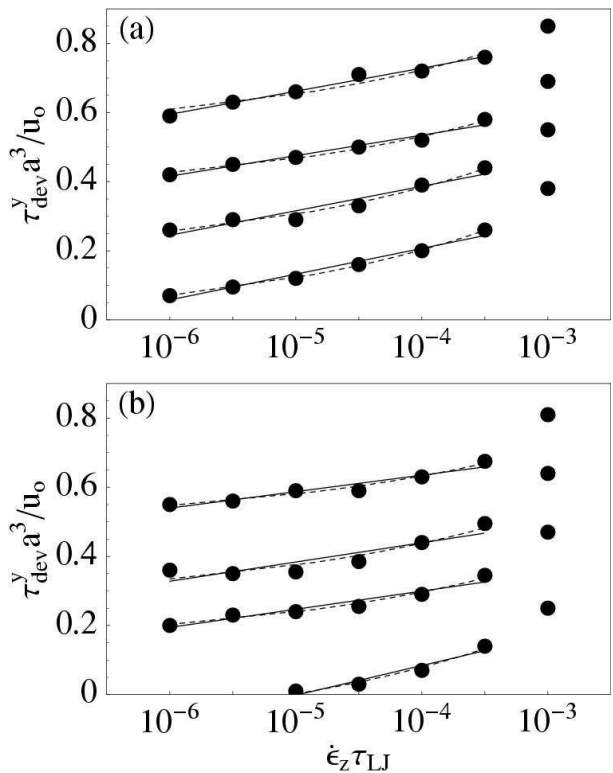}
\caption[Rate dependence of the maximum shear yield stress] {Rate
dependence of the maximum shear yield stress for (a) the polymer glass
and (b) the 80/20 LJ glass. The temperature decreases from
$T=0.3\,u_0/k_B$ (bottom data points) to $T=0.01\,u_0/k_B$ (top data
points) with intermediate values of $T=0.2\,u_0/k_B$ and
$T=0.1\,u_0/k_B$. Also shown are fits to a logarithmic rate dependence
$\tau_{\rm dev}^y=\tau_0+s\ln{\dot{\epsilon}}$ (solid) and a power law
$\tau_{\rm dev}^y=\tau_0+r\dot{\epsilon}^n$ (dashed) with $n=0.2$. Fit
values of $s$ were (a) $0.028\pm 0.003$ and (b) $0.022 \pm 0.02$. }
\label{maxratedep-fig}
\end{center}
\end{figure}

Fig.~\ref{maxratedep-fig} summarizes values for $\tau_{\rm dev}^y$
obtained from the maximum of curves such as those in
Fig.~\ref{stresstraindiffrates-fig} as a function of strain rate. Four
different temperatures were studied, with the highest value,
$T=0.3\,u_0/k_B$, close to $T_g$ and $T=0.01\,u_0/k_B$ far away. This
covers a much wider range than usually explored in experiments.  For a
given temperature and most strain rates, both models give nearly
straight lines in a semilogarithmic plot. The salient observation that
can be made in this figure is that all curves are nearly parallel, and
temperature merely changes the offset value on the stress axis.

\begin{figure}[t]
\begin{center}
\includegraphics[width=8cm]{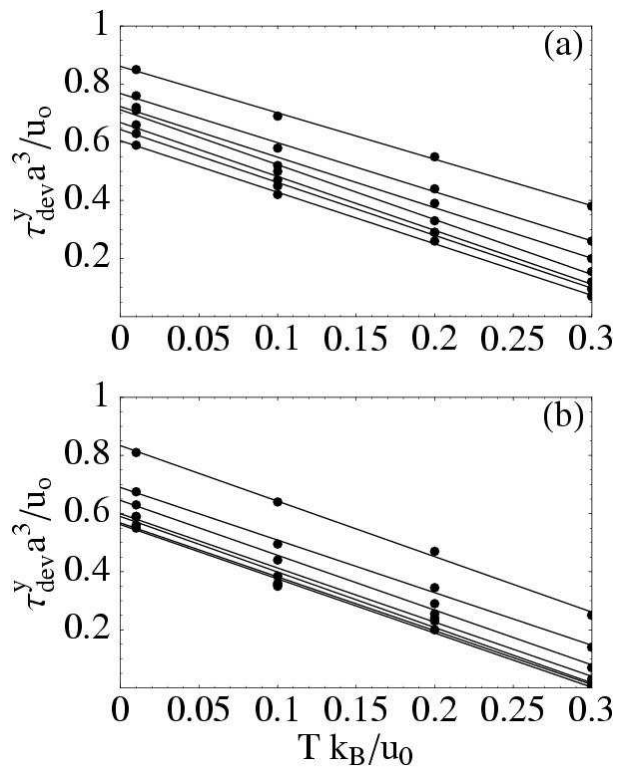}
\caption[Variation of the shear yield stress with temperature]
{Variation of the shear yield stress with temperature for (a) the
polymer glass and (b) the 80/20 LJ glass. The rates vary between
$10^{-3}\tau_{\rm LJ}^{-1}$ and $10^{-6}\tau_{\rm LJ}^{-1}$ as in
Fig.~\ref{maxratedep-fig} (from top to bottom). Linear fits to the
data points for a given rate are also shown.}
\label{tempdep-fig}
\end{center}
\end{figure}

As suggested by the Eyring-model, an obvious way to describe the rate
dependence is through a logarithmic fit of the form $\tau_{\rm
dev}^y=\tau_0+s\ln{\dot{\epsilon}}$ (solid lines).  The observation of
parallel curves then implies a nearly constant prefactor $s$ in front
of the logarithm, and the variation with temperature is described by
$\tau_0$. Often the rate dependence of the shear yield stress of
complex materials is also described with a power-law added to a
constant, i.~e.~ $\tau_{\rm dev}^y=\tau_0+r\dot{\epsilon}^n$ (dashed
lines). As can be seen in Fig.~\ref{maxratedep-fig}, such fits provide
an equally good description of the data.  Due to the small variations
of $\tau_{\rm dev}^y$, a determination of the exponent $n$ via best
fits is very unreliable. Since the curves for different $T$ are nearly
parallel, there is no reason to expect $n$ to vary. The curves in
Fig.~\ref{maxratedep-fig} are for $n=0.2$, which gave the smallest
variation of the prefactor $r$ ($<10\%$) with temperature, but other
choices between 0.1 and 0.3 are also possible. Since $\tau_0>0$ for
$T\leq 0.2u_0/k_B$, the data is not consistent with a pure power law.

Neither functional form provides a good fit to results at the highest
strain rate $10^{-3}\tau_{\rm LJ}^{\-1}$.  We have already shown that
the elastic behavior changes at this shear rate because stress can not
equilibrate throughout the system.  The plastic response will also be
affected by the dynamics of stress distribution, and new behavior is
expected to set in at this rate. Therefore, these points were not
included in the fits shown in Fig.~\ref{maxratedep-fig}.  Note that
experiments are always at much lower strain rates, while most
simulations of shear have been done at $\dot{\epsilon_i} = 10^{-5}$ to
$10^{-1}\tau_{\rm LJ}^{-1}$. A clear distinction between a logarithm
and a power-law dependence would thus only be possible by measuring
the yield stress at lower rates, but unfortunately the simulation time
becomes prohibitively large.

The fact that the curves of Fig.~\ref{maxratedep-fig} are parallel
implies that replotting the shear yield stress as a function of
temperature also leads to parallel behavior. Fig.~\ref{tempdep-fig}
shows that the trend of $\tau_{\rm dev}^y$ with $T$ is in fact linear,
and the slope is nearly rate independent. The rate merely changes
the offset value.

In order to investigate whether the rate dependence is in any way
influenced by the methodology, we have expanded our studies and
considered other model parameters, different thermostat methods and
loading states. In particular, we obtained data analogous to that
shown in Fig.~\ref{maxratedep-fig} for a longer range of the
Lennard-Jones potential $r_c=2.2$a, Nos\'e-Hoover and Langevin
thermostats with different rates, and uniaxial strain as opposed to
the volume-conserving shear in Fig.~\ref{stresstraindiffrates-fig}. We
also considered a system that was quenched into the glassy state 10
times faster than the other systems to investigate cooling rate
effects.  All these different situations show essentially the same
robust scenario: the rate dependence is nearly independent of
temperature with a constant offset that decreases linearly as $T$
increases.  The slope $s$ is generally unaffected within the noise,
but is roughly twice as large for uniaxial strain, and slightly larger
for rapidly quenched initial states.  Again, power-law fits could also
be used to describe the data, and a fixed exponent suffices to
describe all temperatures.

\subsection{Steady shear}
\begin{figure}[b]
\begin{center}
\includegraphics[width=8cm]{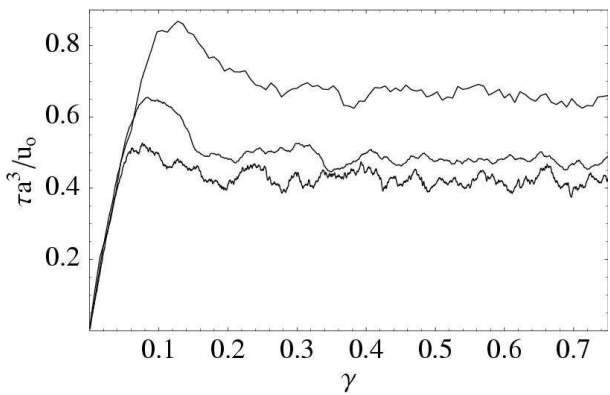}
\caption[Stress strain curves for simple shear]
{Stress strain curves for simple shear of the binary LJ solid at
$T=0.1\,u_0/k_B$ for three different shear rates
$\dot{\gamma}=0.01\tau_{\rm LJ}^{-1},
\dot{\gamma}=0.001\tau_{\rm LJ}^{-1}$ and
$\dot{\gamma}=0.0001\tau_{\rm LJ}^{-1}$ (top to bottom).}
\label{stressstrainsteady-fig}
\end{center}
\end{figure}

Most rheological measurements of soft glassy materials focus on the
steady state stress and not on the initial transient maximum of the
stress-strain curve.  Thus it is interesting to compare the rate
dependence for steady state shear to the above results.  The steady
state can not be studied with the entangled bead-spring model used
above \footnote{The bead-spring model could be used when the chain
length is very short.}, because the shear stresses at accessible
strain rates would soon break covalent bonds.  However previous
studies with short chains \cite{Baljon} and the binary system
\cite{Yamamoto1998,Lacks2001,Barrat2000} show similar trends with
decreasing $T$.  At high $T$ ($\geq 0.7 u_0/k_B$ for the binary
system), Newtonian behavior is observed up to the highest practical
shear rates.  At lower temperatures, where the liquid is in a
supercooled state, there is a crossover from Newtonian behavior at low
rates to power law shear thinning $\sigma \sim \dot{\gamma}^n$ at high
rates.  The crossover rate goes to zero as $T$ decreases to $T_g$, and
at lower temperatures the shear stress is nearly independent of strain
rate \cite{Baljon}.

We focus here on the binary (80/20) LJ system at $T\leq 0.3u_0/k_B$,
varying $T$ from just above $T_g$ to well below.  As described in
Section \ref{model-sec}, shear was imposed by confining the system
between walls separated by $h$ along the z-direction and translating
one at a fixed speed $v_x$. Fig.~\ref{stressstrainsteady-fig} shows
typical behavior of the stress as a function of the average shear
strain $\gamma$.  There is an initial maximum at $\gamma\sim 0.1$ that
corresponds to the transient behavior studied in the previous section.
The stress then decreases rapidly to a steady state plateau
value. This strain softening becomes more pronounced with decreasing
temperature.

\begin{figure}[t]
\begin{center}
\includegraphics[width=8cm]{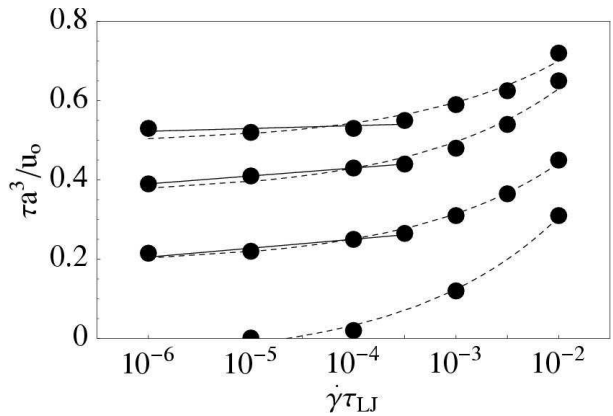}
\caption[Flow curves of the binary LJ glass at different temperatures]
{Rate dependence of the steady state shear stress for the 80/20 LJ
glass at 4 temperatures $T=0.3\,u_0/k_B$ (bottom), $T=0.2\,u_0/k_B$,
$T=0.1\,u_0/k_B$,and $T=0.01\,u_0/k_B$ (top). Also shown are
fits to a logarithmic rate dependence $\tau_{\rm
dev}=\tau_0+s\ln{\dot{\epsilon}}$ (solid) and a power law $\tau_{\rm
dev}=\tau_0+r\dot{\epsilon}^n$ (dashed), where $n=0.3$ for all curves.}
\label{steadystress-fig}
\end{center}
\end{figure}

The average value from the steady state region is plotted vs. shear
rate in Fig.~\ref{steadystress-fig}. At $T=0.3 u_0/k_B$ (lowest
curve), the solid exhibits glassy behavior at high rates, but is
Newtonian at lower rates, which indicates that the temperature is
still above $T_g$. For all temperatures below $T_g$ the curves are
nearly parallel, as for the results for the transient maximum shear
stress in Fig.~\ref{maxratedep-fig}. A logarithmic rate dependence may
be fitted at small rates, but a power-law added to a constant clearly
provides a much better fit over the whole range of rates. The fits for
the temperatures $T\leq 0.2\,u_0/k_B$ shown in
Fig.~\ref{steadystress-fig} use $n=0.3$, for which the prefactor $r$
to the power-law deviates from unity by less than 10\%.  Our results
for the glassy state are very similar to observations reported by
Varnik {\it et al.}  \cite{Varnik2002}, who used the same model with a
larger cutoff value $r_c=2.5\,a$ for the LJ potential and only
considered $T=0.2u_0/k_B$.

The rate dependence of the steady shear stress is very similar to that
of the yield stress (Fig.~\ref{maxratedep-fig}).  In both cases, there
is a rapid rise above the logarithmic fits at rates of
$10^{-3}\tau_{\rm LJ}$ and above.  Although in
Fig.~\ref{steadystress-fig} the system has had time to reach a steady
state, one may wonder whether there is still a change in behavior near
$\dot{\gamma}=10^{-3}\tau_{\rm LJ}^{-1}$.  At higher rates the stress
may not relax between the local yield events discussed below, leading
to a more rapid rise in mean stress.  The results were not extended to
higher shear rates because the temperature in unthermostatted
directions begins to rise slightly above the set temperature even at
$\dot{\gamma}=10^{-2}\tau_{\rm LJ}^{-1}$.

\begin{figure}[t]
\begin{center}
\includegraphics[width=8cm]{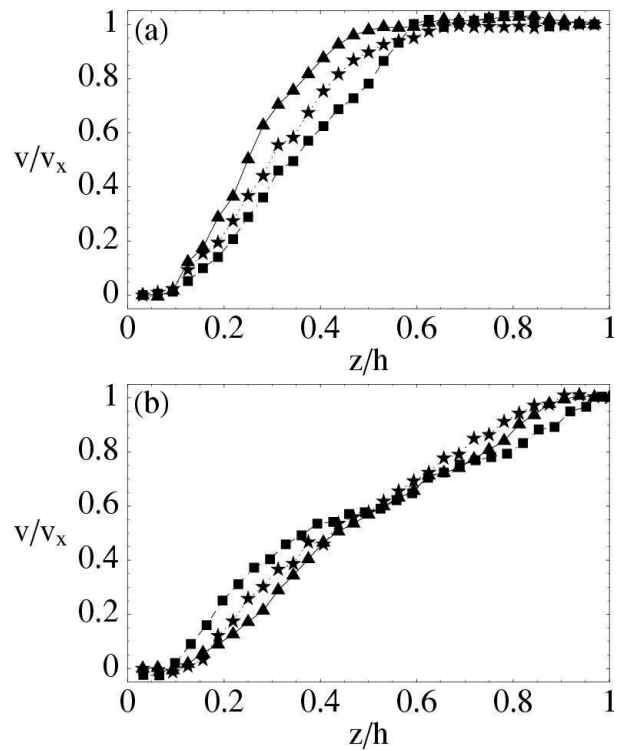}
\vspace{0.1cm}
\caption[Velocity profiles in steady state shear]
{Rescaled velocity profiles for 3 different shear rates
$\dot{\gamma}=10^{-3}\tau_{\rm LJ}^{-1}$ ($\blacktriangle$),
$\dot{\gamma}=10^{-4}\tau_{\rm LJ}^{-1}$ ($\star$),
$\dot{\gamma}=10^{-5}\tau_{\rm LJ}^{-1}$ ($\blacksquare$) at (a)
$T=0.01 u_0/k_B$ and (b) $T=0.2 u_0/k_B$. The velocities are rescaled
by the wall velocity $v_x$ and the positions by the total separation
$h$ of the walls.}
\label{flowprof-fig}
\end{center}
\end{figure}

A plot of the shear stress vs. rate as in Fig.~\ref{steadystress-fig}
assumes that the local strain rate is equal to the average implied by
$v_x/h$.  Fig.~\ref{flowprof-fig} shows the velocity profile for two
different temperatures and several different shear rates. The first
few layers always move with the walls due to the strong coupling.  At
$T=0.2\, u_0/k_B$, the profile in the center region is homogeneous and
here the assumption that $v_s/h$ equals the shear rate is
satisfied. However, at $T=0.01\, u_0/k_B$ shear is localized in the
lower 60\% of the simulation cell, and the upper part of the glass
moves at the constant wall speed. This is a clear indication of shear
banding. The local shear rate is now larger than the average value,
but since the change is only by about a factor of two, the data points
in Fig.~\ref{stressstrainsteady-fig} are not dramatically affected.

In their simulations at $T=0.2\, u_0/k_B$, Varnik {\it et al.}
\cite{Varnik2002} also found a transition from homogeneous flow to
shear banding as the shear rate dropped below $10^{-3}\,\tau_{\rm
LJ}^{-1}$. In our simulations, shear is homogeneous at this
temperature, but the shorter cutoff in our model implies a slightly
lower $T_g$. The increasing amount of shear localization with
decreasing temperature is consistent with a stress peak that becomes
more pronounced as T is lowered (see
Fig.~\ref{stressstrainsteady-fig}). The region with negative slope on
the stress-strain curve signals a mechanical instability and shear
localization. This localization is inhibited in the simulations that
were used to study the transient stress maximum, since they used
periodic boundary conditions in all directions.  Varnik {\it et al.}
\cite{Varnik2002} noted that simulations with Lees-Edwards boundary
conditions \cite{Berthier2002} also suppressed shear banding, but that
the shear stresses obtained from simulations with walls and periodic
boundary conditions are similar.
\subsection{Analysis of the dynamics of the local stress distribution}
\label{jumpsec}

One of the great advantages of molecular simulations is the ability to
reach beyond a measurement of the macroscopic response function and in
addition obtain information about the local dynamics and
rearrangements in nonequilibrium situations. Such information is
essential to construct a clear picture of the underlying microscopic
processes.

Here, we follow this route by decomposing the total simulation cell
into small volume elements of size $7-8 a^3$ and measuring the local
stress tensor in those regions. Since the density is close to $1
a^{-3}$, these regions contain typically 7-8 particles. A particle
number of that size should be sufficient to constitute a locally
transforming region as envisioned in several of the theoretical models
described in Section \ref{theo-sec}.  Some of the models suggested
that local rearrangements replace thermal noise as the source for
activation over barriers, which motivates a study of the stress
changes experienced by local regions.

Our results for general stress states \cite{Rottler2001} have shown
that the deviatoric shear stress $\tau_{\rm dev}$ is the relevant
stress tensor invariant that describes shear deformation. We therefore
calculate the local change $\Delta\sigma_{ij}$ in the stress tensor in
every volume element during a small time interval and then study the
size distribution of changes in the scalar variable $\Delta\tau_{\rm
dev}$ calculated from $\Delta\sigma_{ij}$ (note that $\Delta\tau_{\rm
dev}$ is always positive, since it refers to the deviatioric stress of
$\Delta\sigma_{ij}$) . In general, one expects $\Delta\tau_{\rm dev}$
to fluctuate even in the undriven case, and one should expect to find
a stationary distribution of stress jumps over not too long
timescales. Since the glass is out of equilibrium, it is by
construction not stationary and will exhibit aging phenomena,
etc. Such long timescales, however, are not explored here.

\begin{figure}[t]
\begin{center}
\includegraphics[width=8cm]{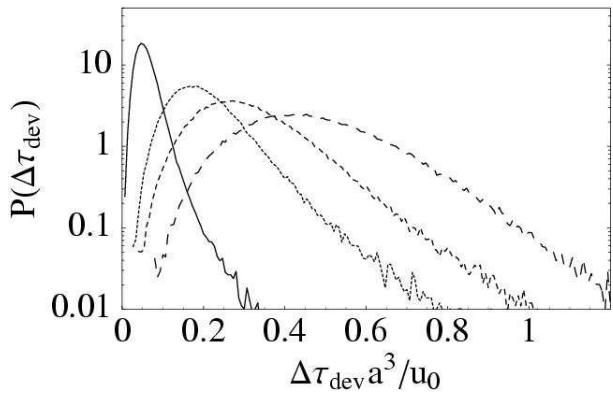}
\caption[Size distribution of jumps $\Delta\tau_{\rm dev}$ for the {\em
unstrained} binary LJ glass] {Size distribution of jumps
$\Delta\tau_{\rm dev}$ for the {\em unstrained} binary LJ glass at
four temperatures $T=0.01 u_0/k_B$ (solid line), $T=0.1
u_0/k_B$,$T=0.2 u_0/k_B$ and $T=0.3 u_0/k_B$ (long dashed line, $T$
increases from left to right). $\Delta\tau_{\rm dev}$ was calculated
for a time difference of $7.5\tau_{\rm LJ}$. The distributions are
stationary over $\sim 10^5
\tau_{\rm LJ}$.}
\label{backgroundjumps-fig}
\end{center}
\end{figure}

Fig.~\ref{backgroundjumps-fig} shows an example of such stationary
distributions for four different temperatures. The jumps correspond to
an average change in local deviatoric stress in small regions over a
time difference of $7.5\tau_{\rm LJ}$. The narrowest distribution is
found at the lowest temperature $T=0.01 u_0/k_B$ and the distribution
widens as the temperature increases to $T=0.3 u_0/k_B$. The cause for
the stress jumps is obviously thermal motion of the particles, and
only small excursions about their positions can occur in the glass
(cage effect).  The part of the distribution at small values of
$\Delta\tau_{\rm dev}$ can be fitted to a Gaussian, but deviations
become visible at large values.

\begin{figure}[t]
\begin{center}
\includegraphics[width=8cm]{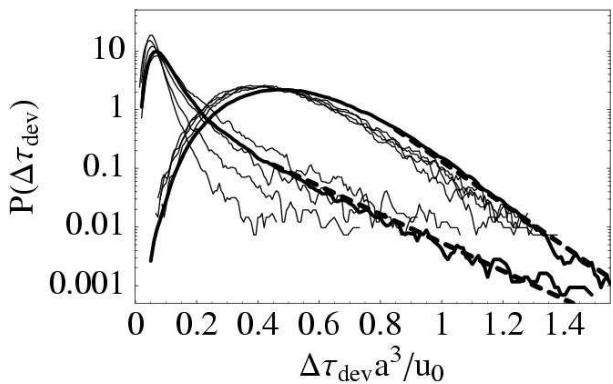}
\caption[Size distribution of jumps in $\tau_{\rm dev}$ for the binary
LJ glass {\em sheared at $\dot{\gamma}=10^{-4}\tau_{\rm LJ}^{-1}$}.]
{Size distribution of jumps in $\tau_{\rm dev}$ for the binary LJ
glass {\em sheared at $\dot{\gamma}=10^{-4}\tau_{\rm LJ}^{-1}$} for
two temperatures $T=0.01 u_0/k_B$ and $T=0.3 u_0/k_B$. Different
curves belong to strains of 0\%, 2.5\%, 4.5\% and 7\% (see
Fig.~\ref{stresstraindiffrates-fig}). Also shown as thick solid lines
are the steady state distributions at the same rate as well as
exponential fits (dashed lines) to the tail of those distributions.}
\label{drivenjumps-fig}
\end{center}
\end{figure}

We are now able to determine how the distribution of stress jumps in
the undriven system changes when the system is macroscopically
strained. Fig.~\ref{drivenjumps-fig} shows several distributions at
the highest ($T=0.3\,u_0/k_B$) and lowest ($T=0.01\,u_0/k_B$)
temperatures considered.  $\Delta\tau_{\rm dev}$ was calculated for
the same time difference $7.5\tau_{\rm LJ}$. For each temperature,
four distributions are shown, which correspond to four different strains
between zero and the maximum shear yield strain. Also shown for
comparison are the distributions in steady shear.

Several observations can be made in this figure. Note first that at
$T=0.3 u_0/k_B$, the unstrained distribution is not changed
dramatically under external driving. Only the tail at large jumps gets
modified. This temperature is above $T_g$ and the shear rate is low
enough that the deviation from Newtonian behavior is small.  This is
in sharp contrast to the situation at $T=0.01 u_0/k_B$, where strain
produces a dramatic increase in the number of large jumps.  The tail
of the distribution can be fit to an exponential form
$\exp(-\Delta\tau_{\rm dev}/\tau_c)$ that extends to larger stresses
as the strain increases to the peak in the stress-strain curve.  The
steady state stress results are close to those at a transient strain
of 4.5\%.  The steady state probability curves decay with the
characteristic stresses $\tau_c=0.18 u_0/a^3$ $(T=0.01 u_0/k_B)$ and
$\tau_c=0.12 u_0/a^3$ $(T=0.3 u_0/k_B)$.

\section{Interpretation and comparison to models of plasticity}
\label{disc-sec}
The above results for the strain rate dependence of shear yielding
offer an opportunity to test the theoretical models described in
Section \ref{theo-sec}. The Eyring model Eq.~(\ref{eyring-eq})
predicts a logarithmic dependence of $\tau^y_{\rm dev}$, and the
prefactor $s$ is given by $k_BT/V^\star$. Since we found typical
values of $s$ between 0.02 and 0.03 for all temperatures, our result
can only be reconciled with the Eyring model if one allows for huge
variations of the ``activation volume'' $V^\star$ between $0.3\,a^3$
and $10\,a^3$. $V^\star$ is a phenomenological fit parameter, but is
typically interpreted as a characteristic volume for a local shear
event. It should then be at least of order the volume per particle,
i.e. of order $a^3$ or larger.  The implied linear change in $V^\star$
with $T$ would also be inconsistent with the observed linear
temperature dependence of $\tau^y_{\rm dev}$. From
Eq.~(\ref{eyring-eq}) the stress would vary as $E/V^\star\sim E/T$,
which is inconsistent with the data.  Besides, the usefulness of the
Eyring expression is reduced if $V^\star$ is allowed to be
temperature dependent.

Modern theories offer a different interpretation of the rate
dependence. Lema\^\i tre's extension \cite{Lemaitre2001} of the STZ
theory \cite{STZ} to include free volume relaxation produces complex
rate dependence that could describe our results.  However, the issue
of temperature dependence has not been addressed in this approach and
we have not examined the issue of free volume relaxation here.

A pure power-law shear thinning behavior was predicted in the
calculations of Berthier {\it et al.} \cite{Berthier2000}. Starting
from very different considerations, these authors found that the
exponent $n$ should have a generic value of 1/3 at a transition
temperature $T_c$. Below $T_c$, the exponent should in principle be
temperature dependent, but their numerical results indicated only a
very weak dependence. Both results are consistent with our findings
near $T_g$, but not at lower temperatures where we find a constant
plus power-law behavior.

Below the glass transition temperature, the SGR model \cite{SGR}
predicts such constant plus power-law behavior of the flow curve with
an exponent of the form $1-x$, where $x$ is an effective noise
temperature in units where the glass transition temperature
$x_g=1$. In steady shear, we found good agreement with this functional
form and an exponent $n=0.3\pm 0.1$. The authors of the SGR model go
to great length in providing an interpretation for $x$. They argue
that in a sheared state, the energy for surmounting a barrier for
rearrangement is not only provided by the thermal energy, but also by
the energy released from rearrangements elsewhere in the
material. This energy diffuses through the material and provides an
effective thermal bath in a mean field sense. The energy released in
such a rearrangement must therefore be of order of the typical yield
energies, which implies that $x$ should be of order unity. When the
yield energies are much larger than typical thermal energies, $x$ will
be independent of the true thermodynamic temperature $T$. Our finding
of an exponent independent of $T$ could thus be rationalized in this
framework.

Our analysis of the distribution of local stress jumps could lend
additional support to the concepts behind the SGR model. Changes in
the local deviatoric stress as shown in Fig.~\ref{drivenjumps-fig} can
activate yield events. We found that the distribution of stress jumps
could be fitted to an exponential distribution with a characteristic
decay stress that varied less than 30\% as $T$ changed from $0.3\,
u_0/k_B$ to $0.01\, u_0/k_B$.  At small values of $\Delta \tau
_{dev}$, the distribution retains the equilibrium form. This result
suggests that the drive generates internal dynamics on a common scale
despite very different thermodynamic temperatures and might provide a
more microscopic justification of the background ``effective noise
temperature'' proposed in the SGR model.

The shear energy released by a yielding local region should indeed
trigger additional yield events at other locations in the
material. However, the ensuing dynamics may be more complicated than
suggested by mean-field approaches.  Many models of material breakdown
include load redistribution mechanisms. In fiber bundle models
\cite{Kloster1997}, for instance, one finds avalanche behavior that
precedes total failure. The avalanche size distribution follows a
power law. When shearing a material, however, a yielded region will
typically reemerge with a different yield energy as assumed in the SGR
model. In ref.~\cite{Zapperi1997}, the simpler situation of an elastic
network with random breaking thresholds of the links was subjected to
an external drive. Damaged elements were replaced with new ones with a
different breaking threshold. Power-law behavior in the size and
duration of failure events was found. Recent work on molecular
\cite{Gagnon2001} as well as simplified models \cite{Baret2002} shows
that load redistribution in the yielding glass mediated by elastic
interactions between rearranging regions can lead to avalanche
behavior.

\section{Conclusions}

The (transient) maximum deviatoric shear stresses for two model
glasses, binary LJ mixtures and polymers, were shown to exhibit
strikingly similar trends with rate and temperature.  The rate
dependence was remarkably constant from $\sim T_g$ to temperatures 30
times lower.  The entire flow curve shifted linearly to lower stress
as $T$ increased.  At low rates, the rate dependence of the peak
stress could be described by a logarithm or constant plus power-law,
where the prefactor and/or exponent did not vary with temperature.  A
more rapid rise in stress was observed when strain rate was increased
to $10^{-3}\tau_{\rm LJ}^{-1}$.  Deviations in the elastic response
also set in at this strain rate, indicating that stress could not
relax throughout the system.  This is not surprising given that the
time to strain to 1\% is comparable to that of sound propagation back
and forth across the system.

The stress for steady shear flow was calculated for the LJ mixture.
Curves for different temperatures were also nearly parallel, shifting
rigidly to lower stresses with increasing temperature. Below strain
rates of $10^{-3}\tau_{\rm LJ}$, the dependence of the flow stress on
rate could be described by a logarithm.  The entire flow curve could
be fit by a constant plus power law with a temperature independent
exponent $n=0.3\pm 0.1$.  However, as for the transient stress, the
stress begins to rise more rapidly at $10^{-3}\tau_{\rm LJ}^{-1}$.  It
is interesting to ask whether the emergence of a power-law
vs. logarithmic behavior is related to the overlap of shear rate and
stress equilibration timescales.  Although the system is in steady
state, regions of a few atoms undergo a substantial yield event after
of order 1\% strain increments.  At strain rates of $10^{-3}\tau_{\rm
LJ}^{-1}$ and above, the time between these yield events will be too
short for stress relaxation in our system.  This point should be kept
in mind when comparing molecular simulations to experiments at lower
strain rates.

The Eyring model was shown to be incompatible even with logarithmic
fits to the shear stress at low rates.  It predicts that the prefactor
of the logarithm should scale as $kT/V^\star$, while the observed
prefactor is nearly independent of temperature.  Fixing this
discrepancy by requiring $V^\star$ to scale linearly with $T$ leads to
unphysically small values of $V^\star$ and is inconsistent with the
linear drop in yield stress with temperature at a given rate.  The
Eyring model has been very helpful in analyzing experimental data, but
typically over a narrow range of temperatures.  It would be
interesting to extend these measurements to very low temperatures and
to higher shear rates. Note that typical room temperature experimental
values for $V^\star$ in polymers correspond to 3 or 4 repeat units
\cite{Ward1983}, which is of the same order as the values we find at
$T=0.1$ or $0.2 u_0/k_B$.

The insensitivity of the rate dependence to temperature changes
indicates that thermal activation is not dominant at the rates
studied. Analysis of the individual particle trajectories offers an
explanation for this.  Even in the limit of zero temperature we find
that the trajectories are exponentially sensitive to changes in rate
and other parameters.  The system does not travel along a single path
through the energy landscape at different rates, but gets deflected
between many possible paths by small perturbations.

Both the STZ and SGR models describe complex dynamics in systems where
temperature is unimportant.  Instead, activation is due to another
internal state variable, either $v_f$ or $x$, that couples to the
external drive via a feedback mechanism.  The original STZ theory
gives a simple linear rise in stress with rate above the yield stress
\cite{STZ}.  However recent generalizations \cite{Lemaitre2001} can
produce more complex rate dependence like that found here.  The SGR
model predicts a constant plus power-law behavior that is also
consistent with our simulations and can account for a temperature
independent exponent $n$.  It remains to be seen whether the
generating mechanism of structural disorder (SGR) or free volume
dynamics (STZ) provides a more useful description of glassy dynamics.

An external drive introduces a new timescale into the problem that
couples to the structural rearrangements. It will therefore modify the
associated relaxation time scales and alter the unperturbed glassy
dynamics, e.g to stop aging and induce rejuvenation \cite{Liu2001}. At
zero temperature, this timescale should be the only one present in the
system, while other competing timescales will arise at finite
temperature. Associated with this timescale is the concept of an
effective temperature \cite{Barrat2000,Ono2001} that also appears in
the SGR model. While this picture is clearly a useful starting point,
the analysis of the jump distribution has shown that the internal
dynamics of the flowing glassy solid is much more intricate and
strongly calls for extensions of analytical models beyond mean-field
concepts. Such a treatment should describe more accurately the load
redistribution mechanisms and the propagation of released shear energy
on a microscopic level. A preliminary analysis has shown that local
plastic events occur in avalanches over a wide range of length and
time scales.

Finally, we note that the above analytic models of glassy rheology and
viscoplasticity are all scalar models that cannot describe
self-organization on larger length scales such as shear
bands. Tensorial versions of the STZ theory are being discussed to
address banding and necking \cite{Langer2001}. These are promising
approaches that should continue to benefit from insight gained from
molecular simulations, in particular the load redistribution
mechanisms of shear yielding zones.

\section{Acknowledgements}
We thank J.~S.~Langer and M.~L.~Falk for useful discussions.
Financial support from the Semiconductor Research Corporation (SRC)
and NSF grant No. DMR0083286 is gratefully acknowledged. The
simulations were performed with {\em LAMMPS 2001} \cite{Lammps2001}, a
molecular dynamics package developed by Sandia National Laboratories.


\begin{thebibliography}{99}
\bibitem{Larson1999} R.~G.~Larson, {\em The structure and rheology of
complex fluids} (Oxford University Press, New York 1999).
\bibitem{Rottler2001} J.~Rottler, M.~O.~Robbins, Phys.~Rev.~E {\bf
64}, 051801 (2001).
\bibitem{STZ}M.~L.~Falk and J.~S.~Langer, Phys.~Rev. E {\bf 57}, 1971
  (1998), M.~L.~Falk and J.~S.~Langer, M.~R.~S. Bulletin {\bf 25}, 40
  (2000).
\bibitem{Lemaitre2001} A.~Lema\^\i tre, Phys.~Rev.~Lett. {\bf 89},
195503 (2002) and cond-mat/0206260.
\bibitem{Hasan1995} O.~A.~Hasan and M.~C.~Boyce, Polymer Eng.~and
Sci.~{\bf 35}, 331 (1995).
\bibitem{Liu2001} A.~J.~Liu and S.~R.~Nagel (Eds.), {\it
Jamming and Rheology}, (Taylor \& Francis, London, 2001)
\bibitem{Barrat2000} J.~L.~Barrat, L.~Berthier, Phys.~Rev.~E {\bf 63},
012503 (2000), L.~Berthier, J.~L.~Barrat, Phys.~Rev.~Lett {\bf 89},
095702 (2002).
\bibitem{Ward1983} I.~M.~Ward, {\em Mechanical Properties of Solid
Polymers} (John Wiley \& Sons, New York 1983).
\bibitem{Eyring1936} H.~Eyring, J.~Chem.~Phys. {\bf 4}, 283 (1936).
\bibitem{SGR} P.~Sollich, F.~Lequeux, P.~H\'ebraud, M.~E.~Cates,
  Phys.~Rev.~Lett. {\bf 78}, 2020 (1997), P.~Sollich, Phys.~Rev. E
  {\bf 58}, 738 (1998), S.~Fielding, P.~Sollich, M.~E.~Cates,
  J.~Rheol. {\bf 44} 323 (2000).
\bibitem{Berthier2000} L.~Berthier, J.-L.~Barrat, J.~Kurchan,
Phys.~Rev.~B {\bf 61}, 5464 (2000).
\bibitem{Monthus1996} C.~Monthus and J.-P.~Bouchaud, J.~Phys.~A:
Math.~Gen.~{\bf 29}, 3847 (1996).
\bibitem{tc-comm} $T_c$ refers to a dynamical transition temperature,
where the relaxation time in the model diverges as a power law.
\bibitem{Kremer1990} K.~Kremer and G.~S.~Grest, J.~Chem.~Phys. {\bf
92} 5057 (1990).
\bibitem{Puetz2000} M.~P\"utz, K.~Kremer, G.~S.~Grest,
Europhys.~Lett. {\bf 49}, 735 (2000).
\bibitem{Kob1995} W.~Kob and H.~C.~Andersen, Phys.~Rev.~E {\bf 51},
4626 (1995), Phys.~Rev.~E {\bf 52}, 4134 (1995).
\bibitem{Kob1997} W.~Kob and J.-L.~Barrat, Phys.~Rev.~Lett.~{\bf 78},
4581 (1997).
\bibitem{Kob1997b} W.~Kob, C.~Donati, S.~J.~Plimpton, P.~H.~Poole, and
S.~C.~Glotzer, Phys.~Rev.~Lett.~{\bf 79}, 2827 (1997).
\bibitem{Bennemann1998} C.~Bennemann, W.~Paul, K.~Binder and
B.~D\"unweg, Phys.~Rev.~E {\bf 57} 843 (1998).
\bibitem{Allen1987} M.~P.~Allen, D.~J.~Tildesley, {\em Computer Simulations of Liquids} (Oxford University Press, Oxford 1987).
\bibitem{Thomson1990}P.~A.~Thompson and M.~O.~Robbins, Phys.~Rev.~A
{\bf 41}, 6830 (1990).
\bibitem{Voight1928} W. Voight, Lehrbuch der Kristallphysik (Teubner,
Leipzig, 1928).
\bibitem{Baljon} A.~R.~C.~Baljon and M.~O.~Robbins, Science {\bf 271},
482 (1996); A.~R.~C.~Baljon and M.~O.~Robbins, ``Stick-Slip Motion,
Transient Behavior, and Memory in Confined Films,'' in {\em
Micro/Nanotribology and its Applications}, edited by B.~Bhushan
(Kluwer, Amsterdam, 1997) pp. 533-553.
\bibitem{Yamamoto1998} R.~Yamamoto and A.~Onuki, Phys.~Rev.~E {\bf
58}, 3515 (1998).
\bibitem{Lacks2001} D.~J.~Lacks, Phys.~Rev.~Lett. {\bf 87}, 225502
(2001).
\bibitem{Varnik2002} F.~Varnik, L.~Bocquet, J.-L.~Barrat, L.~Berthier,
Phys.~Rev.~Lett. {\bf 90}, 095702 (2003). 
\bibitem{Berthier2002} L.~Berthier and J.-L.~Barrat,
J.~Chem.~Phys. {\bf 116}, 6228 (2002).
\bibitem{Kloster1997} M.~Kloster, A.~Hansen, P.~C.~Hemmer,
Phys.~Rev.~E {\bf 56}, 2615 (1997)
\bibitem{Zapperi1997} S.~Zapperi,A.~Vespignani, H.~E.~Stanley, Nature
{\bf 388}, 658 (1997)
\bibitem{Gagnon2001} G.~Gagnon, J.~Patton, and D.~J.~Lacks,
Phys.~Rev.~E {\bf 64}, 051508 (2001)
\bibitem{Baret2002} J.-C.~Baret, D.~Vandembroucq, and S.~Roux, e-print cond-mat/0206523.
\bibitem{Ono2001} Ian K.~Ono, Corey S.~O'Hern, D.~J.~Durian, Stephen
A.~Langer, Andrea J.~Liu, and Sidney R.~Nagel, Phys.~Rev.~Lett.~{\bf
89}, 095703 (2002).
\bibitem{Langer2001} J.~S.~Langer, Phys.~Rev. E {\bf 64}, 011504 (2001),
 L.~O.~Eastgate, J.~S.~Langer, L.~Pechenik, e-print cond-mat/0206363.
\bibitem{Lammps2001} http://www.cs.sandia.gov/$\sim$sjplimp/lammps.html.
\end{thebibliography}
\end{document}